\documentclass[11pt]{article}
 \def\bb{\bibitem} \def\lb{\label}
\def\be{\begin{equation}} \def\ee{\end{equation}}
\def\ba{\begin{eqnarray}} \def\ea{\end{eqnarray}} \def\part{\partial}

\begin{document}
\begin{titlepage}

\title{
The black holes of topologically massive gravity}
\author{Karim Ait Moussa$^{a,b}$ \thanks{Email:
karim.aitmoussa@wissal.dz}, G\'erard
Cl\'ement$^a$ \thanks{Email: gclement@lapp.in2p3.fr} and C\'edric
Leygnac$^a$ \thanks{Email: leygnac@lapp.in2p3.fr}  \\ \\ {$^a$ \small
Laboratoire de  Physique Th\'eorique LAPTH (CNRS),} \\ {\small
B.P.110, F-74941 Annecy-le-Vieux cedex, France}\\
{$^b$ \small Laboratoire de Physique Math\'ematique et Physique
Subatomique,}\\{\small D\'epartement de Physique, Facult\'e des
Sciences, Universit\'e Mentouri,}\\{\small Constantine 25000, Algeria}}
\maketitle
\begin{abstract}
We show that an analytical continuation of the Vuorio solution to
three-dimensional topologically massive gravity leads to a
two-parameter family of black hole solutions, which are geodesically
complete and causally regular within a certain parameter range. No
observers can remain static in these spacetimes. We discuss their
global structure, and evaluate their mass, angular momentum, and
entropy, which satisfy a slightly modified form of the first law of
thermodynamics.
\end{abstract}
\end{titlepage}
\setcounter{page}{2}

It is well known that Einstein gravity in 2+1 dimensions is
dynamically trivial, without propagating degrees of freedom. The
addition to the Einstein-Hilbert action of a gravitational
Chern-Simons term leads to topologically massive gravity \cite{djt}
(TMG), with massive spin 2 excitations. A long standing question
\cite{deser} is that of the existence of Schwarzschild- or Kerr-like
black hole solutions to TMG. The equations of TMG with a negative
cosmological  constant are trivially solved \cite{kaloper} by the BTZ
black-hole  metric \cite{btz}. Other black hole solutions to
cosmological TMG were  written down in \cite{nutku,gurses}, however it is
possible to show that these solutions are not bona fide black holes
for the range of parameters considered in \cite{nutku,gurses}. A number of
exact solutions to TMG with vanishing cosmological constant are known
\cite{vuo,HMP,NB,part}, but these do not include black hole solutions.
The purpose of this Letter is to show that a simple extension of the
Vuorio solution \cite{vuo} leads to regular TMG black holes, and to
present an introductory exploration of their properties.

The field equations of TMG are
\be\lb{tmg}
G^{\mu}_{\;\;\nu}+\frac{1}{\mu} \, C^{\mu}_{\;\;\nu} = 0,
\ee
where $G^{\mu}_{\;\;\nu} \equiv R^{\mu}_{\;\;\nu}-
\frac{1}{2} \,R\,\delta^{\mu}_{\;\;\nu}$ is the Einstein
tensor,
\be
C^{\mu}_{\;\;\nu} \equiv
\varepsilon^{\mu\alpha\beta}\,D_{\alpha}\,(R_{\beta\nu}-
\frac{1}{4}\,g_{\beta\nu}\,R)
\ee
is the Cotton tensor (the antisymmetrical tensor is
$\varepsilon^{\mu\alpha\beta} = |g|^{-1/2}\eta^{\mu\alpha\beta}$, with
$\eta^{012} = +1$), and
$\mu$ is the topological mass constant. Vuorio searched for
stationary rotationally symmetric solutions to
these equations, and noticed that they could be solved exactly by
assuming a constant $g_{tt}$, which he normalized to the Minkowski
value\footnote{Contrary to Vuorio, we use the signature ($- + +$)
for a Lorentzian metric.} $g_{tt} = -1$. Vuorio's solution is,
in units such that $\mu = +3$,
\be\lb{vu}
ds^2 = -\bigg[d{\tilde t} - (2\cosh\sigma + \tilde{\omega})d\tilde{\varphi}\bigg]^2 +
d\sigma^2 + \sinh^2\sigma\,d\tilde{\varphi}^2,
\ee
where $\tilde{\varphi}$ is assumed to be periodic with period $2\pi$,
and we have reintroduced an integration constant $\tilde{\omega}$
which Vuorio set to $-2$ for regularity. This spacetime is
homogeneous, with constant curvature scalars, and admits four Killing
vectors generating the Lie algebra of $SL(2,R) \times U(1)$
\cite{vuo,ortiz}. However, it has one undesirable property. From
(\ref{vu}) we obtain $g_{\tilde{\varphi}\tilde{\varphi}} =
-3\cosh^2\sigma + 4\tilde{\omega}\cosh\sigma - \tilde{\omega}^2 - 1$,
which is negative definite if $\tilde{\omega}^2 < 3$, and negative
outside a critical radius $\sigma_c$ if $\tilde{\omega}^2 > 3$, leading
to closed timelike circles\footnote{The same is true for the purported
black-hole solutions of \cite{nutku} with $\Lambda = 0$, $2J > M$, or
of \cite{gurses} with $\lambda = 0$, $a_0 > 0$.} for all $\sigma > \sigma_c$.

To overcome this indesirable violation of causality, let us
analytically continue the solution (\ref{vu}) by the combined imaginary
coordinate transformation (which does not change the overall
Lorentzian signature) $\tilde{t} = i\sqrt{3}t$, $\tilde{\varphi} =
i(\rho_0/\sqrt{3})\varphi$, with $\rho_0$ an arbitrary positive
constant, leading to
\be\lb{bh1}
ds^2 = 3\bigg[dt - \bigg(\frac{2\rho}3+\omega\bigg)d\varphi\bigg]^2
+ \frac{d\rho^2}{\rho^2-\rho_0^2} -
\frac{\rho^2-\rho_0^2}3\,d\varphi^2,
\ee
where $\omega = \rho_0\tilde{\omega}/3$, the new radial coordinate
is $\rho = \rho_0\cosh\sigma$, and $\varphi$ is again assumed to be an
angular variable with period $2\pi$. This new solution of the field
equations can also be derived directly \`a la Vuorio by making the
unconventional ansatz $g_{tt} = +3$ (instead of $-1$). The fact that
the Killing vector field $\part_t$ is spacelike, not timelike, means
that there can be no static observers in such a geometry. Furthermore
it is easily seen that no linear combination of the Killing vectors
$\part_t$ and $\part_{\varphi}$ can remain timelike for $\rho \to
\infty$. This situation is quite similar to that inside the
ergosphere of the Kerr metric, except that here the ergosphere extends
to spacelike infinity. As in the case of the Kerr ergosphere, the
solution is however locally stationary. At a given radius $\rho$, observers
moving with uniform angular velocity $\Omega$ remain timelike,
$(\part_t + \Omega\part_{\varphi})^2 < 0$, provided
\be
\frac{2\rho + 3\omega - \sqrt{\rho^2-\rho_0^2}}{r^2} < \Omega <
\frac{2\rho + 3\omega + \sqrt{\rho^2-\rho_0^2}}{r^2}.
\ee

The divergence of the metric component $g_{\rho\rho}$ at $\rho =
\pm\rho_0$ suggests that (\ref{bh1}) is a black
hole solution, which is made obvious by rearranging the metric as
\be\lb{bh2}
ds^2 = -\frac{\rho^2-\rho_0^2}{r^2}\,dt^2 +
\frac{d\rho^2}{\rho^2-\rho_0^2}  + r^2\bigg(d\varphi -
\frac{2\rho + 3\omega}{r^2}\,dt\bigg)^2,
\ee
with
\be\lb{r2}
r^2 = \rho^2 + 4\omega\rho + 3\omega^2 + \rho_0^2/3.
\ee
While this metric is not asymptotically Minkowskian, the squared lapse
$N^2 = (\rho^2 - \rho_0^2)/r^2$ and the shift $N^{\varphi} = -(2\rho +
3\omega)/r^2$ respectively go to 1 and 0 at spacelike infinity $\rho
\to \pm\infty$.
For $\rho_0^2 > 0$, there are two horizons located at
$\rho_{h\pm} = \pm\rho_0$, of perimeter and angular velocity
\be
A_{h\pm} \equiv 2\pi r_{h\pm} = 2\pi|2\rho_0 \pm 3\omega|/\sqrt{3}, \qquad
\Omega_{h\pm} = 3/(\pm2\rho_0+3\omega). \lb{hor}
\ee
If $\omega \ne \mp 2\rho_0/3$, the metric may be extended through these
horizons by the usual Kruskal method. Just as in the case of the BTZ
black holes, the metric (\ref{bh1}) is regular for all $\rho \neq
\pm\rho_0$, so that the maximally extended spacetime is geodesically
complete, with a Penrose diagram similar to that of the Kerr black
hole (except of course for the ring singularity). However,
singularities in the causal structure do occur for
a range of values of $\omega$. If $\omega^2 < \rho_0^2/3$, the
Killing vector $\part_{\varphi}$ is everywhere spacelike. On the
contrary, if $\omega^2 > \rho_0^2/3$, $\part_{\varphi}$ becomes timelike
in the range $\rho_{c-} < \rho < \rho_{c+}$, with
\be
\rho_{c\pm} = -2\omega \pm\sqrt{\omega^2-\rho_0^2/3}\,.
\ee
It is easily seen that these two zeros of
$(\part_{\varphi})^2$ are timelike lines belonging to
the same stationary Kruskal patch, so that the acausal regions $r^2 <
0$ are safely hidden behind the horizon for an observer at $\rho = +\infty$
if $\omega > 0$, which we shall assume henceforth. The Penrose diagram
for the maximally extended spacetime with the acausal regions cut out
is the same as for
Reissner-Nordstr\"{o}m black holes. The limiting case $\rho_0
= 0$ leads to extreme black holes, with a double horizon at $\rho =
0$. In this case, there is an acausal region
$(\part_{\varphi})^2 < 0$ behind the horizon for
all positive values of $\omega$. The resulting Penrose
diagram (again with the acausal regions cut out) is identical to that
of extreme Reissner-Nordstr\"{o}m black holes.

The case $\omega = 2\rho_0/3$ deserves special
consideration. In this case the metric (\ref{bh2}) reduces to
\be\lb{sch}
ds^2 = -\frac{\rho-\rho_0}{\rho +5\rho_0/3}\,dt^2 +
\frac{d\rho^2}{\rho^2-\rho_0^2} + (\rho
+5\rho_0/3)(\rho+\rho_0)\bigg(d\varphi - \frac{2dt}{\rho
+5\rho_0/3}\bigg)^2.
\ee
This has only one horizon at $\rho = +\rho_0$, where Kruskal extension
can be carried out as usual. Near the causal singularity
$\rho = -\rho_0$, the metric (\ref{sch}) can be approximated by
\be
ds^2 \simeq 3\,dt^2 + d\sigma^2, \quad d\sigma^2 = -
\frac{d\rho^2}{2\rho_0(\rho+\rho_0)} +
(2\rho_0/3)(\rho+\rho_0)\,d\hat{\varphi}^2,
\ee
with $d\hat{\varphi} = d\varphi - 3\,dt$. Clearly the two-dimensional
metric $d\sigma^2$ becomes null for $\rho \to -\rho_0$, and is
non-extendible because of the periodicity condition on $\varphi$, so
that $\rho = -\rho_0$ is a spacelike singularity of the metric
(\ref{sch}). The corresponding Penrose diagram is thus identical to
that of the Schwarzschild black hole.

The even more special case $\omega = \rho_0= 0$ lies at the intersection
of the preceding case and of the extreme black hole case $\rho_0 =
0$. In this case the metric (\ref{bh2}) reduces to
\be\lb{vac}
ds^2 = -dt^2 + \frac{d\rho^2}{\rho^2} +
\rho^2\bigg(d\varphi-\frac2{\rho}\,dt\bigg)^2,
\ee
which is devoid of horizons, and thus qualifies as the ground state or
``vacuum'' of our two-parameter family of black-hole solutions. This
metric does not appear to be extendible beyond the manifest
singularity $\rho = 0$, which can be shown to be at finite affine
distance. Moreover, the angular velocity of stationary observers
approaching this singularity increases without bound, so that the very
concept of a Penrose diagram breaks down.

The black hole metric (\ref{bh2}) depends on two parameters $\rho_0$
and $\omega$, which should somehow be related to the physical
parameters, mass and angular momentum. The standard approach for
computing these quantities in the case of non-asymptotically flat
spacetimes\footnote{The computation of energy and angular momentum
for asymptotically Minkowkian or asymptotically (A)dS solutions to
TMG recently carried out in \cite{DT03} cannot be used here.} uses
the idea of quasilocal energy \cite{BY}. From the action functional
for a self-gravitating system with boundary conditions on a given
hypersurface, one derives canonically a Hamiltonian, given by the sum
of a bulk integral, which vanishes on shell, and of a surface term.
The quasilocal energy is the (substracted) on-shell value of the
Hamiltonian in the limit where the spatial boundary is taken to
infinity. A canonical formulation of topologically massive gravity
was given in \cite{dexi}. However integrations by part were freely
performed in \cite{dexi}, so that at present we do not know what is
the correct surface term. Instead we shall bypass the standard
quasilocal approach by suitably extending the recently proposed super
angular momentum approach to the computation of conserved quantities
in 2+1 gravity \cite{black}. This approach relies on the observation
that the dimensional reduction of a self-gravitating system with two
Killing vectors $\part_t$ and $\part_{\varphi}$ leads to a mechanical
system with the $SL(2,R) \sim SO(2,1)$ invariance. This mechanical
system has a conserved super angular momentum, two components of
which may be identified as the mass and angular momentum of the
(2+1)-dimensional gravitating configuration. These identifications
have been shown in \cite{black} to correspond to a well-defined
``finite part'' prescription for computing the (unsubstracted)
quasilocal conserved quantities, and to lead to consistent results
for Einstein-scalar and (up to some gauge ambiguity) for Einstein-Maxwell
black holes.

The conserved super angular momentum for TMG has been given in
\cite{part}. The general stationary rotationally symmetric metric
may be written in the 2+1 form
\be\lb{21}
ds^2=\lambda_{ab}(\rho) \, dx^a \, dx^b + \zeta^{-
2} (\rho) \, R^{-2}(\rho) \, d\rho^2 ,
\ee
where $\lambda$ is the $2 \times 2$ matrix
\be\lb{la}
\lambda = \left( \begin{array}{cc}
   T+X & Y \\
    Y & T-X \end{array} \right),
\ee
$R^2 = {\bf X}^2 =-T^2+X^2+Y^2$ is the Minkowski pseudo-norm of the
vector ${\bf X}(\rho)=(T,X,Y)$, and the function $\zeta(\rho)$ allows
for arbitrary reparametrizations of the radial coordinate $\rho$. In
the gauge $\zeta= 1$, the conserved generalized angular momentum for TMG is
\be\lb{superam}
{\bf J} = \frac1{2\kappa}\bigg({\bf X}\wedge{\bf X}' +
\frac1{2\mu}\bigg[{\bf X}'\wedge({\bf X}\wedge{\bf X}') -
2{\bf X}\wedge({\bf X}\wedge{\bf X}'')\bigg]\bigg),
\ee
where $\kappa = 8\pi G$ is the Einstein gravitational constant, the prime
is the derivative $d/d\rho$, and
the wedge product is defined by $({\bf X} \wedge {\bf Y})^A =$ \linebreak
$\eta^{AB}\epsilon_{BCD}X^C Y^D$ (with $\eta^{AB}$ the inverse
Minkowski metric, and $\epsilon_{012} = +1$). Assuming that the
identifications of Einstein-scalar or Einstein-Maxwell black hole
conserved quantities proposed in \cite{black} can be extended to TMG,
the (2+1)-dimensional mass and angular momentum are given by
\ba
M & = & -2\pi\,J^Y \,, \lb{M}\\
J & = &  2\pi(J^T-J^X) \,. \lb{J}
\ea

We first test these formulas on the example of the BTZ solution, for
which the quasilocal mass and angular momentum have recently been
computed in the wider framework of a Poincar\'e gauge theory
\cite{GHHM} (the action for this theory reduces to that of TMG when
a constraint for vanishing torsion is added). The BTZ metric is given by
\ba
ds^2 & = & (-2l^{-2}\rho + {\rm M}/2)\,dt^2 - {\rm J}\,dt\,d\varphi
+ (2\rho + {\rm M}l^2/2)\,d\varphi^2 \nonumber \\ & & + [4l^{-2}\rho^2 -
({\rm M}^2l^2-{\rm J}^2)/4]^{-1}\,d\rho^2\,,
\ea
with $\Lambda = -l^{-2}$ the cosmological constant. The parametrization
(\ref{la}) of this metric corresponds to \cite{black}
\be
{\bf X} = \left|\begin{array}{l}
(1-l^{-2})\rho + (1+l^{-2}){\rm M}l^2/4\\
-(1+l^{-2})\rho - (1-l^{-2}){\rm M}l^2/4\\
-{\rm J}/2 \end{array}\right.\,,
\ee
with $\zeta = 1$. The computation of the superangular momentum
(\ref{superam}) is straightforward and gives the TMG mass and angular
momentum of the BTZ black holes in terms of the Einstein conserved
quantities ${\rm M}$ and  ${\rm J}$,
\be\lb{MJBTZ}
M = \frac{\pi}{\kappa}\bigg({\rm M} - \frac{{\rm J}}{\mu l^2}\bigg)\,,
\quad J = \frac{\pi}{\kappa}\bigg({\rm J}- \frac{{\rm M}}{\mu}\bigg)\,.
\ee
These values (which reduce to the Einstein values in the limit $\mu
\to \infty$) coincide with the values (22) and (23) obtained in
\cite{GHHM} in the special case of a vanishing torsion ${\cal T} = 0$
(the identification is $2\ell\theta_L = -1/\mu$, $\Lambda_{eff} =
-\Lambda = l^{-2}$, $\ell = \kappa = \pi$, $\chi = 1$).

Strengthened by this agreement, we proceed with the computation of the
conserved quantities $M$ and $J$ for the $\Lambda = 0$ TMG black
holes. The parametrization (\ref{la}) of the metric (\ref{bh2})
corresponds to
\be\lb{bh3}
{\bf X} = \left|\begin{array}{l}
\rho^2/2 + 2\omega\rho + 3(\omega^2+1)/2 + \rho_0^2/6\\
-\rho^2/2 - 2\omega\rho - 3(\omega^2-1)/2 - \rho_0^2/6\\
-2\rho - 3\omega \end{array}\right.\,,
\ee
with $\zeta =1$. The computation of the super angular momentum
(\ref{superam}) leads to
\be
M = \frac{\pi}{\kappa}\,\omega \,, \quad
J = \frac{\pi}{\kappa}(\omega^2-5\rho_0^2/9) \,. \lb{MJ}
\ee
Surprisingly, the mass depends only on the parameter $\omega$, and not
on the ostensible horizon radius $\rho_0$. These relations can be
inverted to yield
\be
\omega = \frac{\kappa}{\pi}\,M\,, \quad \rho_0^2 =
\frac{9\kappa^2}{5\pi^2}\,(M^2 - \pi J/\kappa)\,.
\ee
Extreme black holes thus have $J = \kappa M^2/\pi$,
Schwarzschild-like black holes have $J = -\kappa M^2/4\pi$, while the
mass and angular momentum vanish for the vacuum solution (\ref{vac}),
as expected.

What are the thermodynamical properties of these black holes? The
Hawking temperature depends only on the metric, not on the particular
theory of gravity considered, and is given by the inverse of the
period in imaginary time,
\be\lb{temp}
T_H \equiv
\frac1{2\pi}\,n^{\rho}\part_{\rho}N\,|_{(\rho = \rho_h)} = \frac{\zeta
RR'}{2\pi\sqrt{V}}\,\bigg|_{(\rho = \rho_h)}\,.  \ee We obtain here
the temperature \be\lb{temp1} T_H =
\frac{\sqrt{3}}{2\pi}\,\frac{\rho_0}{2\rho_0 + 3\omega}\,,
\ee
which
as usual vanishes for extreme black holes ($\rho_0 = 0$). On the
other hand, the black hole entropy should depend on the specific
theory under consideration, and we have no reason to expect that it is
given by the Einstein value $S_E = (2\pi/\kappa)A_h$. To determine the
black hole entropy, we use the first law of thermodynamics, according
to which the entropy variation is given by
\be \frac{\part S}{\part
M}\bigg|_J= T_H^{-1}.  \ee
This can be integrated to yield
\be\lb{S} S
= \frac{2\pi^2}{3\sqrt{3}\kappa}\,(5\rho_0 + 6\omega)\,,  \ee
up to an
arbitrary additive function of $J$. Assuming that this function
vanishes, we obtain from (\ref{S}), (\ref{temp1}), (\ref{MJ}) and
(\ref{hor}) a modified form of the first law of black hole thermodynamics:
\be
dM = T_H\,dS + \frac12\,\Omega_h\,dJ\,.  \ee
The anomalous factor
$1/2$ in front of the angular velocity is another surprising effect
of TMG. Finally, a simple quadratic combination of the same
(undifferentiated) quantities yields the Smarr-like formula
\be M  =
T_H S + \Omega_h J\,, \ee
to be compared with the Smarr-like formula
for 2+1 Einstein-scalar black holes \cite{black} $M  = T_H S/2 +
\Omega_h J$.

A problem with the value (\ref{MJ}) for the black hole mass is that,
contrary to the case of four-dimensional Einstein gravity, the
gravitational constant $\kappa$ must be negative in TMG to avoid the
occurrence of ghosts \cite{djt}. This means that, for $\omega^2 >
\rho_0^2/3$, causally regular black holes, with $\omega > 0$, have a
negative mass (as well as a negative entropy). For $\omega^2 <
\rho_0^2/3$, all black holes are regular, but only those with $\omega
< 0$ have a positive mass (but a negative entropy). We do not know how
to solve this problem, but point out that a similar problem arises
with the BTZ black holes viewed as solutions of TMG with negative
gravitational constant. It follows from (\ref{MJBTZ}) that, for $\kappa
< 0$, regular BTZ black holes, with ${\rm M} > 0$ and ${\rm J}^2 \le {\rm
M}^2l^2$, necessarily have a negative TMG mass $M$ if $\mu^2l^2 > 1$,
and do not necessarily have a positive TMG mass if $\mu^2l^2 \le 1$.

We have shown that an analytical extension of the Vuorio solution to
TMG leads to a two-parameter family of black hole solutions, which are
causally regular within a certain parameter range. While their metric
is not asymptotically Minkowskian, the AdM lapse and shift functions go
to the Minkowski values 1 and 0 at spacelike infinity, so that their
Penrose diagrams are similar to those of Kerr or Reissner-Nordstr\"om
black holes. In this respect, these black hole spacetimes are closer
to four-dimensional black holes than the asymptotically AdS BTZ black
holes of \cite{btz}. No observers can remain static in these spacetimes,
however stationary observers are allowed, which is all that is needed
to discusss physical experiments such as wave scattering. We have
evaluated the mass, angular momentum, and entropy of these black
holes, which satisfy a slightly modified form of the first law of
thermodynamics. At present these evaluations remain tentative. Our
formulas (\ref{M}) and (\ref{J}) for computing the black hole mass and
angular  momentum generalize formulas previously derived and tested in
the case of Einstein-scalar or Einstein-Maxwell black holes, and have
been tested here in the specific case of the BTZ solution to
cosmological TMG. However a full computation of the quasilocal energy
and angular momentum in TMG should be carried out in order to
ascertain the validity of our evaluations (the computations of
\cite{GHHM} cannot be adapted for this purpose because the field
equations of the Poincar\'e gauge theory considered there are stronger
than those of TMG, and do not admit our black holes as solutions). A
direct computation of the black hole entropy is also desirable. We
intend to address these questions, and to further elucidate the
properties of our black holes, in a future publication.

\end{document}